\documentclass[twocolumn,showpacs,preprintnumbers,amssymb]{revtex4}

\usepackage{graphicx}
\usepackage{bm}

\begin{document}
\title{Role of electrostatics in the texture of islands in free standing ferroelectric liquid crystal films}

\author{Jong-Bong Lee$^{1,*}$}
\author{Robert A. Pelcovits$^2$}
\author {Robert B. Meyer$^1$}
\affiliation{$^1$The Martin Fisher School of Physics, Brandeis
University, Waltham, MA 02454\\$^2$Department of Physics, Brown
University, Providence, RI 02912}

\date{\today}
\begin{abstract}
Curved textures of ferroelectric smectic C* liquid crystals
produce space charge when they involve divergence of the
spontaneous polarization field. Impurity ions can partially screen
this space charge, reducing long range interactions to local ones.
Through studies of the textures of islands on very thin
free-standing smectic films, we see evidence of this effect, in
which materials with a large spontaneous polarization have static
structures described by a large effective bend elastic constant.
To address this issue, we calculated the electrostatic free energy
of a free standing film of ferroelectric liquid crystal, showing
how the screened coulomb interaction contributes a term to the
effective bend elastic constant, in the static long wavelength
limit. We report experiments which support the main features of
this model.
\end{abstract}

 \maketitle

The inclusion of sub-micron particles in a very thin free standing
ferroelectric Smectic C (Sm-C*) film induces the growth of islands,
circular regions of thicker material, usually with a central point
topological defect surrounded by a texture which satisfies certain
boundary conditions and minimizes an appropriate free
energy\cite{bob-mclc,jb}. In the simplest case, this is a pure bend
texture, with the molecules oriented tangentially everywhere in the
island, and the topological defect, or disclination, at the center.
This yields a radial spontaneous polarization $\bm{P}=\pm
P_0\bm\hat{r}$ whose divergence produces space
charge\cite{bob_ferro}.  Thus the electrostatic interaction in this
pure bend island should play an important role in the energetics of
the liquid crystal texture.  In our initial studies\cite{jb}, we
performed a purely elastic analysis of the free energy of the
observed textures, and found that our observations of textural
transitions as a function of island diameter were well explained by
a large ratio of bend to splay elastic constants.  We observed that
this large ratio correlated with a large spontaneous polarization in
the material studied.  This suggested that the observed large
effective bend elastic constant actually arose from the
electrostatic interaction.  In this paper, we report experiments and
theory which develop the evidence for this scenario, which explains
well our observations.

The long range electrostatic interaction arising from divergence of
the spontaneous polarization field has been studied previously, as
has the screening of that interaction by dissolved impurity ions in
the liquid crystal\cite{degenne}. In Sm-C* free standing films, the
impurity ion effect first was observed qualitatively by Pindak
\textit{et al.} \cite{pindak1}. Subsequently, the screening effect
by impurity ions in Sm-C* free standing films was studied
experimentally in Ref.~\cite{Mac,dislocation}. Theoretically Okano
showed that in the long wavelength limit in a bulk ferroelectric
liquid crystal, the screened Coulomb interaction reduced to an
effective bend elastic constant, whose magnitude depends on the
concentration of impurity ions and the size of the spontaneous
polarization \cite{okano}. This analysis followed that of Palierne,
who discussed the ionic screening of the electric field produced by
the flexoelectric polarization in the nematic phase in three
dimensions \cite{palierne}. For a thin film, the geometry of
screening effects is different, since the sources of the electric
field and the screening ions are confined to two dimensions, while
the field propagates in three dimensions. Nevertheless, we thought
that such ionic screening of the space charge arising from
spontaneous polarization might change the long range electrostatic
energy into a local effective elastic term even in the case of thin
films.

\begin{figure}
\centering
\includegraphics[width=2.0in]{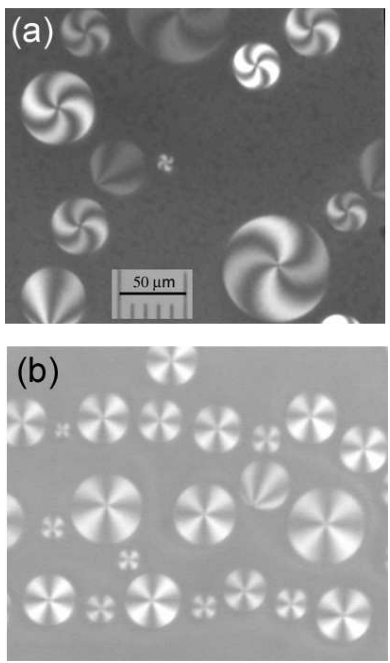}
\caption{Textures of islands viewed between crossed polarizers.
(a) Islands shown in CS1024, typical of a high spontaneous
polarization material (46.7 nC/cm$^2$). (b) Islands seen in
CS1015, a low spontaneous polarization material(6.6
nC/cm$^2$).}\label{fig:islands}
\end{figure}
A textural transformation which we have studied in the islands is
from a pure bend (tangential) pattern to a reversing spiral
texture\cite{jb,kraus,loh}. In the reversing spiral texture, shown
in Fig.~\ref{fig:islands} (a) the tangential boundary conditions
of the c-director (a unit vector parallel to the projection of the
tilted molecular orientation direction on the plane of the film)
still apply at both the outer boundary and the point defect at the
center, but the c-director switches to a radial orientation
between these two limits.  This partially replaces bend with splay
deformation, reducing the elastic energy if the splay elastic
constant is less than the bend elastic constant.  As islands grow,
this transition occurs at a critical island size, which is then a
measure of the relative magnitudes of these two elastic constants.
We investigated the textures of islands in commercial
ferroelectric liquid crystal mixtures, which have a Sm-C* phase at
room temperature. The ferroelectric liquid crystal mixtures we
used are the CS series(CS1015, CS1024, CS2005) by Chisso and the
RO series(RO318, RO322, RO330) by RODIC. Based on the company
specifications, measured at room temperature, CS1024 and CS2005
have relatively high spontaneous polarization(46.7 nC/cm$^2$ and
72.6 nC/cm$^2$, respectively) and CS1015, RO318, RO322, and RO330
are very low spontaneous polarization materials(6.6 nC/cm$^2$, 3
nC/cm$^2$, 3.8 nC/cm$^2$, and 3.8 nC/cm$^2$, respectively).

Islands commonly created in CS1024 and CS1015 are shown in
Fig.~\ref{fig:islands}(a) and (b). In islands on the film with a
high spontaneous polarization, the pure bend texture is rarely seen.
Most of them show a reversing spiral texture or a boojum
texture.(Fig.~\ref{fig:islands}(a)) On the other hand, as shown in
Fig.~\ref{fig:islands}(b), all of the islands are pure bend islands
or boojum islands at equilibrium; RO318, RO322, and RO330 show very
similar textures, in which a reversing spiral island is observed
only at an unusually large island size. We also never observed pure
bend islands in a ferroelectric liquid crystal compound with an
extremely high spontaneous polarization\cite{w399}. Hence the
reversing spiral texture, which is an indicator of a large ratio of
bend to splay elastic constants, is directly correlated with a large
spontaneous polarization.

\begin{figure}
\centering
\includegraphics[width=3.0in]{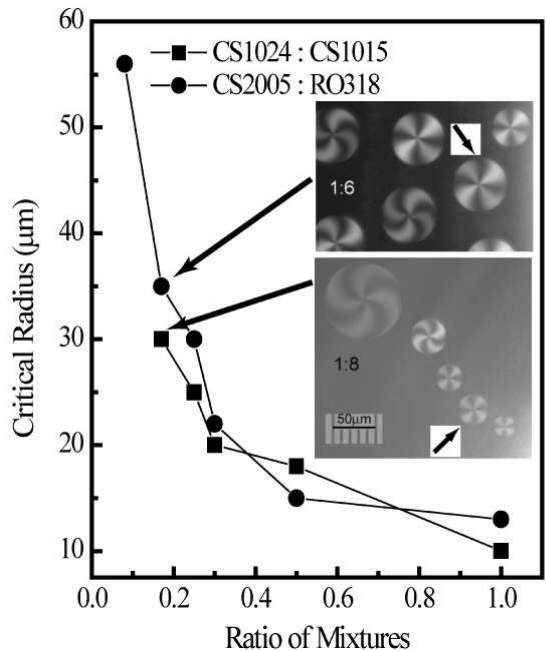}
\caption{Critical radius for transition from pure bend to reversing
spiral vs. weight ratio of a high spontaneous polarization to a
lower spontaneous polarization material. The islands marked by
arrows are just above the critical radius. Top: A mixture of CS1024
and CS1015 with a 1:8 mass ratio. Bottom: A mixture of CS2005 and
RO318 with a 1:6 mass ratio.}\label{fig:ongoing}
\end{figure}
To observe the spontaneous polarization dependence of the textural
transition from a pure bend into a reversing spiral, we investigated
the textures of islands created in mixtures of two ferroelectric
liquid crystals with high and low spontaneous polarizations. The
critical size for a pure bend island to start to change into a
reversing spiral texture was measured as the ratio of the two
materials was varied, as shown in Fig.~\ref{fig:ongoing}, where data
on mixtures of CS1024 and CS1015 and CS2005 and RO318 are shown. The
real time transformation is shown in Ref.\cite{jb} The islands
marked by the arrow are just initiating the transformation to the
reversing spiral texture. Bigger islands than the marked ones have
already transformed into the reversing spiral texture, but smaller
islands are still in the pure bend texture. A higher proportion of
the material with high spontaneous polarization causes a pure bend
island to transform into a reversing spiral texture at a smaller
critical radius. Other mixtures of the high and low spontaneous
polarization materials show very similar behavior.

\begin{table}[h]
\centering \caption{Spontaneous polarization and elastic constants
determined by the electric field quenching experiment.}\label{tab:constants}
\begin{tabular}{cccc}
\hline \hline
Materials &$P_0$(10$^{-5}$esu/cm)  & $K_b$(10$^{-12}$erg)  & $K_s$(10$^{-12}$erg)\\
\hline
CS1024 &7.97 &4.27 &2.43 \\
1:2\footnote{Ratio of CS1024 to CS1015}  &5.49 &3.12 &3.62  \\
1:3  &5.27 &2.09 &2.83  \\
1:5  &2.55 &0.89 &1.12 \\
1:8  &2.45 &0.71 &1.45  \\
CS1015 &2.45 &0.39 &2.64 \\
\hline
\end{tabular}
\end{table}
In order to relate our study of the quasi-static textural
transformation to fundamental liquid crystal properties, we
undertook measurements of fundamental liquid crystal properties of
the mixtures in which we made textural observations.  Knowing that
the screening of the space charges by mobile ions is a slow
process, we undertook measurements of the elastic properties of
these thin films in fast dynamic processes, to obtain values of
``bare'' elastic constants, as well as spontaneous polarizations.
We measured the elastic constants and spontaneous polarizations of
the free standing films (4 smectic layers thick) created by
CS1024, CS1015, and their mixtures by monitoring the change of
intensity of light scattered by c-director fluctuations due to
quenching of the fluctuations by a strong electric field pulse,
following the work of Rosenblatt \textit{et al.}\cite{Rosenblatt}.
The results are listed in Table~\ref{tab:constants}. As we
expected, a larger proportion of the higher spontaneous
polarization material increases the spontaneous polarization of
the mixture. However, the bare splay elastic constants measured
with fast dynamics were found to be larger than the bend elastic
constants in the mixtures. This should favor the pure bend
texture, in disagreement with our observations in the quasi-static
limit, which indicated that the bend elastic constant is larger.
This supported our guess that the screened electrostatic
interactions were contributing to the apparently large
quasi-static bend elastic constant.

To address this issue theoretically, we calculate the
electrostatic free energy of a very thin film. We assume that the
liquid crystal film lies in the plane $z=0$, surrounded by vacuum
on both sides. Mathematically we treat the film as infinitesimally
thin, but introduce the film thickness $a$ in an appropriate
dimensional fashion. Denoting the dielectric constant of the film
(assumed to be spatially uniform) by $\varepsilon$, the dielectric
function for the three--dimensional space is given by:
\begin{equation}
\label{epsilon} \varepsilon^\prime(z)=1+(\varepsilon-1) a \delta(z).
\end{equation}

Gauss' Law then reads:

\begin{equation}
\label{gauss}
\mathbf{\nabla\cdot}[\varepsilon^\prime(z)\bm{E}(\bm{r})]=4\pi\rho(\bm{r}),
\end{equation}
where $\rho(\bm{r})=\rho_{ion}(\bm{r})+\rho_{P}(\bm{r})$ is the sum
of the ionic and ferroelectric space charge densities respectively.
These densities are proportional to $\delta (z)$. Note that $\mathbf
{r}=(\bm{r}_\perp, z)$ and $\bm{\nabla}=(\bm{\nabla}_\perp,\partial/\partial z)$ are three--dimensional vectors.

Using Eq.~(\ref{epsilon}), Gauss' Law can be rewritten as,
\begin{eqnarray}\label{eq:gauss}
\nonumber \bm{\nabla\cdot}[\varepsilon^\prime(z)\bm{E}]&=&
\bm{\nabla\cdot}[(1+(\varepsilon-1) a \delta(z))\bm{E}]\\
\nonumber &=&(1+(\varepsilon-1) a\delta(z))\bm{\nabla\cdot E}\\
&&+(\varepsilon-1) aE_z \delta^\prime (z),
\end{eqnarray}
where $\delta^\prime(z)\equiv {d\delta (z)}/{dz}$. Introducing the
electrostatic scalar potential $\Phi(\bm{r})$ defined as usual by
$\bm{E}=-\nabla\Phi$ we obtain a Poisson--like equation from
Eq.~(\ref{eq:gauss}),
\begin{equation}\label{eq:poisson}
-(1+(\varepsilon-1) a\delta(z))\nabla^2\Phi(\bm{r})-(\varepsilon-1)
a \delta^\prime (z) \frac{\partial \Phi(\bm{r})}{\partial z}=4 \pi
\rho(\bm{r}).
\end{equation}

We now treat the ionic charge density using the Debye--H\"{u}ckel
approximation. Assuming there are \textit{n} species of ionic
impurities, and that the average concentration of the $i^{th}$ type
of impurity is $c_i$, the local concentration $c_i(\bm{r})$ of this
impurity at position $\bm{r}$ is given by the Boltzmann
distribution:
\begin{equation}
c_i(\bm{r})=c_i \exp(-\beta e_i \Phi (\bm{r})) \delta(z),
\end{equation}
where $e_i$ is the charge of the impurity of type \textit{i}, and
$\beta\equiv1/k_BT$. The total ionic charge density at position
$\bm{r}$ is then given by:
\begin{equation}
\rho_{ion}(\bm{r})=\sum_{i=1}^n e_i c_i \exp(-\beta e_i \Phi
(\bm{r}))\delta(z).
\end{equation}

We insert this charge density into Eq.~(\ref{eq:poisson}), and
linearize in $\Phi$ to obtain a Poisson--Boltzmann equation:
\begin{eqnarray}
\label{PB1} \nonumber -(1+(\varepsilon-1)
a\delta(z))\nabla^2\Phi(\bm{r})-(\varepsilon-1) a
\delta^\prime (z) \frac{\partial \Phi(\bm{r})}{\partial z}\\
+4 \pi \beta \sum_{i=1}^n e_i^2 c_i \Phi(\bm{r}) \delta(z) =4 \pi
\rho_P(\bm{r}).
\end{eqnarray}

Replacing the space charge $\rho_P$ by a delta function source, we
obtain the equation obeyed by the Green's function $G(\bm{r})$
associated with Eq.~(\ref{PB1}):
\begin{eqnarray}
\label{PB2} \nonumber -(1+(\varepsilon-1) a\delta(z))\nabla^2
G(\bm{r})-(\varepsilon-1) a
\delta^\prime (z) \frac{\partial G(\bm{r})}{\partial z}\\
+\kappa G(\bm{r}) \delta(z) =4 \pi \delta (\bm{r}).
\end{eqnarray}
where, the inverse two--dimensional (2D) Debye screening length
$\kappa$ is given by:
\begin{equation}
\kappa =4 \pi \beta \sum_{i=1}^n e_i^2 c_i,\label{kappa}
\end{equation}
Fourier transforming Eq.~(\ref{PB2}) we obtain:
\begin{equation}
\label{greeneqn} q^2 G(\bm{q}) + \int \frac{d q_z^\prime}{2 \pi}
G(\bm{q}_\perp, q_z^\prime) ((\varepsilon-1) a q_\perp^2+
\kappa)=4\pi.
\end{equation}
where $\bm{q}_\perp$ is the projection of $\bm{q}$ in the plane of
the film, and we have used the fact that G is an even function of
$q_z$. From Eq.~(\ref{greeneqn}) we find:
\begin{equation}
\label{Gsolution} G(\bm{q})={4 \pi -((\varepsilon-1) a
q^2_\perp+\kappa)\int\frac{dq_z^\prime}{2\pi}G(\bm{q}_\perp,q_z^\prime)
\over q_z^2+q_\perp^2}.
\end{equation}

Integrating both sides of this equation over $q_z$ we
obtain:
\begin{equation}\label{eq:Gq}
\int G(\bm{q}_\perp,q_z){dq_z}={4 \pi^2 \over
q_\perp + {(\varepsilon-1) a  q_\perp^2\over 2}+ {\kappa\over 2}},
\end{equation}
which we will now use to obtain an expression for the screened
electrostatic energy.

The electrostatic energy $F_{elec}$ of the film is given by:
\begin{equation}\label{eq:elec_energy}
F_{elec}=\frac{1}{2}\int{\rho_P(\bm{r}_\perp) \Phi(\bm{r})d^3\bm{r}},
\end{equation}
where the space charge
$\rho_P(\bm{r}_\perp)=-\bm{\nabla\cdot}\bm{P}$. In terms of the Green's function $G(\mathbf{r})$, $F_{elec}$ can be expressed as:

\begin{equation}
\label{energy}
F_{elec}=\frac{1}{2}\int\int d^3r d^3r^\prime \rho_P(\mathbf{r}_\perp) \rho_P(\mathbf{r}_\perp^\prime)\delta(z) \delta(z^\prime) G(\mathbf{r - r^\prime}),
\end{equation}
which in Fourier space yields:
\begin{equation}
F_{elec}=\frac{1}{2}\int {d^2 q_\perp\over (2\pi)^2}\rho_P(\mathbf{q}_\perp)\rho_P(-\mathbf{q}_\perp)\int {dq_z\over 2\pi}G(\mathbf{q}_\perp,q_z).
\end{equation}
Using Eq.~(\ref{eq:Gq}) we obtain:
\begin{equation}
F_{elec}=\frac{1}{2}\int {d^2 q_\perp\over (2\pi)^2}\rho_P(\mathbf{q}_\perp)\rho_P(-\mathbf{q}_\perp){2 \pi \over q_\perp + {(\varepsilon - 1) a  q_\perp^2\over 2}+{\kappa\over 2}}.
\end{equation}

In the long--wavelength limit, where $q_\perp$ is much smaller than the inverse 2D screening length $\kappa$, $F_{elec}$ simplifies to:
\begin{eqnarray}
F_{elec}&=&\frac{2\pi}{\kappa}\int {d^2 q_\perp\over (2\pi)^2}\rho_P(\mathbf{q}_\perp)\rho_P(-\mathbf{q}_\perp)\\
&=&\frac{1}{2}\int\frac{4\pi}{\kappa}({\nabla\cdot}\bm{P})^{2}d
\bm{r}_{\perp}\label{freenergy}.
\end{eqnarray}

The polarization vector is given by
$\bm{P}=P_0\bm{\hat{z}\times\hat{c}}$, where $P_{0}$ is the
magnitude of a spontaneous polarization. Using this expression in
Eq.~(\ref{freenergy}) we find:
\begin{equation}
F_{elec}=\frac{1}{2}\int4\pi\lambda_{2D}P_{0}^{2}(\hat{z}\cdot\nabla
\times\bm{c})^2d \bm{r}_{\perp}.
\end{equation}
where $\lambda_{2D}\equiv 1/\kappa$ is the 2D Debye screening
length.

Hence, the total free energy of the film including the Frank elastic
energy is
\begin{equation}\label{fig:screen_energy}
F=\frac{1}{2}\int[K_{s}(\bm{\nabla\cdot
\hat{c}})^{2}+(K_{b}+4\pi\lambda_{2D}P_{0}^{2})(\bm{\hat z\cdot
\nabla\times \hat{c}})^{2}]d\bm{r}_{\perp}
\end{equation}
where $K_s$ and $K_b$ are the splay and bend elastic constants
respectively. Thus the screening by the impurity ions has rendered
the elastic free energy density local. Therefore, an effective bend
elastic constant can be defined:
\begin{equation}\label{effective}
K_{b}^{eff}=K_{b}+4\pi\lambda_{2D}P_{0}^{2},
\end{equation}
similar to the result obtained for bulk ferroelectric liquid
crystals \cite{okano}.

\begin{figure}
\centering
\includegraphics[width=2.0in]{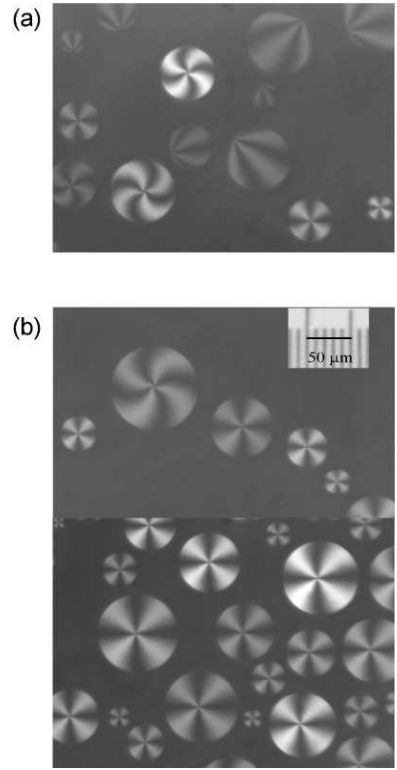}
\caption{Effect of impurity ions. (a). These islands are observed
on a film composed of RO318 and CS2005 with a 1:5 composition
ratio. (b). After doping with an organic salt, most islands are in
the pure bend texture. The critical radius for the transition to
the reversing spiral texture has increased due to the smaller
effective bend elastic constant.}\label{fig:impurity}
\end{figure}

We have shown that the electrostatic energy due to space charge
created by divergence of the spontaneous polarization, screened by
impurity ions dissolved in the liquid crystal, can become a local
effective term in the elastic free energy in the case $q\ll
1/\lambda_{2D}$. This means that a term $\lambda_{2D} P_0^2$ can
dominate the bare bend elastic constant $K_b$ in that limit. This
is reflected in the change of the critical size for the transition
from a pure bend island to a reversing spiral, when varying the
spontaneous polarization or the concentration of the impurity
ions, which tunes $\lambda_{2D}$. The combination of large
spontaneous polarization and low concentration of screening ions
produces the largest increase in the effective bend elastic
constant.

In order to see the effect of the impurity ions, we added free ions
to the material to reduce the 2D Debye screening length as shown in
Eq.~(\ref{kappa}). We doped our ferroelectric liquid crystal
mixtures with an organic salt, tetrabutylammonium bromide(TBAB)
(99$+\%$ purity by Acros Organic) at saturated concentration in a
mixture of RO318 and CS2005 with a composition ratio of 1:5. Before
the salt was added, reversing spiral islands existed at a small size
(Fig.~\ref{fig:impurity}(a)). As shown in the top of
Fig.~\ref{fig:impurity}(b), most of islands on the film doped with
TBAB are in the pure bend texture.  Reversing spiral islands only
appear at a larger critical size than seen in
Fig.~\ref{fig:impurity}(a). Although this result is qualitative, it
indicates the trend of the effect.  It would be desirable to find an
appropriate salt with greater solubility, and to make quantitative
measurements of ionic concentration for this experiment.

In summary, we studied the textural transformation from a pure bend
island to a reversing spiral texture in islands on thin films of
Smectic C* ferroelectric liquid crystals. We observed that the
transition size of the islands depends on the spontaneous
polarization of the material. Based on this observation, we proposed
that the spontaneous polarization and Debye screening length
regulate the effective bend elastic constant in the long wavelength
limit. The typical Debye screening length in these materials has
subsequently been measured to be about 0.7$\mu m$\cite{debye}, in
separate experiments to be reported elsewhere. Our typical island
size is tens of micrometers, so the conditions for use of this
effective bend elastic constant to describe the electrostatic energy
appear to be reasonable. As a qualitative test of this model, we
observed that added impurity ions reduced the effective bend elastic
constant.

\section*{Acknowledgments}
We thanks Namjea Woo for the gift of the RO series of ferroelectric
mixtures. RAP was supported in part by the NSF under grant no.
DMR--0131573. This research was supported at Brandeis University by
NSF grant no. DMR--0322530.\\

$^*$ present address: Department of Physics, POSTECH, Kyungbuk
790-784, Republic of Korea

\end{document}